\documentclass{emulateapj}

\begin{document}

\shortauthors{Morrow et al.}
\shorttitle{Brown Dwarf Disks in TW Hya}

\title{Spitzer IRS Observations of Disks around Brown Dwarfs in the TW Hydra
Association\altaffilmark{1}}

\author{
A. L. Morrow\altaffilmark{2},
K. L. Luhman\altaffilmark{2,3},
C. Espaillat\altaffilmark{4},
P. D'Alessio\altaffilmark{5},
L. Adame\altaffilmark{6},
N. Calvet\altaffilmark{4},
W. J. Forrest\altaffilmark{7},
B. Sargent\altaffilmark{7},
L. Hartmann\altaffilmark{4},
D. M. Watson\altaffilmark{7},
\& C. J. Bohac\altaffilmark{7}
}

\altaffiltext{1}{
Based on observations made with the {\it Spitzer Space Telescope}, which is
operated by the Jet Propulsion Laboratory at the California Institute of
Technology under NASA contract 1407.}

\altaffiltext{2}{Department of Astronomy and Astrophysics,
The Pennsylvania State University, University Park, PA 16802;
morrow@astro.psu.edu.}

\altaffiltext{3}{Visiting Astronomer at the Infrared Telescope Facility, which
is operated by the University of Hawaii under Cooperative Agreement no. NCC
5-538 with the National Aeronautics and Space Administration, Office of Space
Science, Planetary Astronomy Program.}

\altaffiltext{4}{Department of Astronomy, The University of Michigan,
500 Church Street, 830 Dennison Building, Ann Arbor, Michigan 48109.}

\altaffiltext{5}{Centro de Radioastronom\'\i a y Astrof\'\i sica, UNAM,
Apartado Postal 72-3 (Xangari), 58089, Morelia, Michoac\'an, M\'exico.}

\altaffiltext{6}{Instituto de Astronom\'\i a, UNAM, Apartado Postal 70-264,
Ciudad Universitaria, 04510, M\'exico DF, M\'exico.}

\altaffiltext{7}{Department of Physics and Astronomy,
The University of Rochester, Rochester, NY 14627.}

\begin{abstract}

Using SpeX at the NASA Infrared Telescope Facility and 
the {\it Spitzer} Infrared Spectrograph, we have obtained 
infrared spectra from 0.7 to 30~\micron\ for three young brown dwarfs 
in the TW Hydra Association ($\tau\sim10$~Myr), 2MASSW~J1207334-393254, 
2MASSW~J1139511-315921, and SSSPM~J1102-3431.
The spectral energy distribution for 2MASSW~J1139511-315921 is consistent
with a stellar photosphere for the entire wavelength range of our data 
while the other two objects exhibit significant excess emission at 
$\lambda>5$~\micron.
We are able to reproduce the excess emission from each brown dwarf using 
our models of irradiated accretion disks. According to our model fits,
both disks have experienced a high degree of dust settling. 
We also find that silicate emission at 10 and 20~\micron\ is absent from the 
spectra of these disks, indicating that grains in the upper disk layers
have grown to sizes larger than $\sim5$~\micron.
Both of these characteristics are consistent with previous observations 
of decreasing silicate emission with lower stellar masses and older ages.
These trends suggest that either 1) the growth of dust grains, and perhaps 
planetesimal formation, occurs faster in disks around brown dwarfs 
than in disks around stars, or 2)
the radii of the mid-IR-emitting regions of disks are smaller for
brown dwarfs than for stars, and grains grow faster at smaller disk radii.
Finally, we note the possible detection of an unexplained emission feature near
14~\micron\ in the spectra of both of the disk-bearing brown dwarfs.

\end{abstract}


\section{Introduction}
\label{sec:intro}

One of the first steps toward the formation of planets is the growth and 
settling of dust grains in a circumstellar accretion disk \citep{wei93}.
A common diagnostic of grain evolution in 
disks is the silicate emission near 10~\micron\ \citep{nat00}. 
Measurements of this feature have been performed for young
stars ($\tau\sim1$~Myr) across a wide range of spectral types, from 
early B to late M \citep{fur05,kes06,sic07}. In these data, silicate emission 
becomes weaker at later spectral types, which could be explained by more 
advanced grain evolution or sedimentation in disks around low-mass objects 
or a luminosity dependence of the radius in the disk at which silicate 
emission is produced \citep{kes07,sic07}.

Recent studies also have begun to explore the evolution of silicate emission
with time for low-mass systems.
Observations with the {\it Spitzer Space Telescope} have found that
disks around low-mass stars and brown dwarfs exhibit weaker silicate emission 
in Upper Scorpius \citep[$\tau\sim5$~Myr,][]{sch07} than in Chamaeleon~I 
\citep[$\tau\sim3$~Myr,][]{apa05} and Taurus \citep[$\tau\sim1$~Myr,][]{fur05},
which is consistent with the growth of grains to larger sizes, the
settling of dust to midplane, or both, as time goes on.
However, according to a recent study, this correlation between silicate 
emission and age does not apply to the brown dwarf 2MASSW~J1207334-393254 
\cite[henceforth 2M~1207-3932,][]{giz02} in the TW Hya Association 
\citep[TWA, $\tau\sim10$~Myr,][]{mam05,bar06}. \citet{ria07} constructed a
mid-infrared (IR) spectral energy distribution (SED) for this object using 
broad-band photometry from \citet{ste04} and \citet{ria06}, which seemed
to indicate the presence of silicate emission. As a result, they concluded
that the disk around 2M~1207-3932 has experienced little dust
processing relative to stars at the same age. 

To investigate the evolution of silicate emission in brown dwarf disks more 
definitively, we have performed mid-IR spectroscopy on 2M~1207-3932 and 
two other brown dwarfs in TWA with the {\it Spitzer} Infrared 
Spectrograph \citep[IRS;][]{hou04}. 
In this Letter, we describe these observations and supporting near-IR 
spectroscopy (\S~\ref{sec:obs}), examine these data for silicate and 
continuum emission from circumstellar dust (\S~\ref{sec:exc}), 
and fit these data with the predictions of disk models (\S~\ref{sec:model}).

\section{Observations}
\label{sec:obs}

\subsection{Near-infrared Spectroscopy}
\label{sec:spex}

We obtained low-resolution near-IR spectra of 2M~1207-3932 and 
two other brown dwarfs in TWA, 
2MASSW~J1139511-315921 \citep[2M~1139-3159,][]{giz02} and SSSPM~J1102-3431 
\citep[SS~1102-3431,][]{sch05a}, with SpeX \citep{ray03} at the NASA Infrared 
Telescope Facility (IRTF) on the night of 2005 December 14. 
These data were reduced with the Spextool package \citep{cus04} and 
corrected for telluric absorption \citep{vac03}. The final spectra extend 
from 0.8-2.5~\micron\ and exhibit a resolving power of $R=100$.
We flux calibrated the spectra using photometry at $J$, $H$, and $K_s$
from the Point Source Catalog of the Two-Micron All-Sky Survey 
\citep[2MASS,][]{skr06}.
To measure spectral types from these data, we compared them to SpeX
data for young objects that have been classified at optical wavelengths.
The strengths of the TiO, VO, and H$_2$O absorption bands indicate a spectral
type of M8.5 for each object, which is consistent with the 
optical types of M8-M8.5 reported by \citet{giz02} and \citet{sch05a}.
The spectra of 2M~1139-3159 and SS~1102-3431 are slightly redder than
the spectrum of 2M~1207-3932 in a manner that is consistent with reddening
by interstellar dust. Therefore, we dereddened the spectra of 2M~1139-3159 
and SS~1102-3431 by $A_V=0.8$ and $A_V=0.5$, respectively, to match the data 
for 2M~1207-3932. 

\subsection{Mid-infrared Spectroscopy}
\label{sec:irs}

We obtained low-resolution mid-IR spectra of 2M~1139-3159, 2M~1207-3932, 
and SS~1102-343 on 3 July 2005, 29 July 2006, and 3 July 2005, respectively, 
with IRS aboard {\it Spitzer} as a part of the Guaranteed Time Observations of
the IRS instrument team. We used both low-resolution IRS modules, 
Short-Low and Long-Low, which cover 5.3-14 and 14-40~$\mu$m, 
respectively, with a resolution of $\lambda/\Delta \lambda \sim90$.
The total exposure time for each target was $\sim4000$~sec.
The spectra were extracted and calibrated from the basic calibrated data 
using the standard SMART data analysis package for IRS \citep{hig04}.  
The spectra were reduced with the methods that have been previously applied
to IRS data for low-mass members of Taurus \citep{fur05}.

\section{Analysis}
\label{sec:analysis}

\subsection{Disk Emission}
\label{sec:exc}

\begin{figure}
\plotone{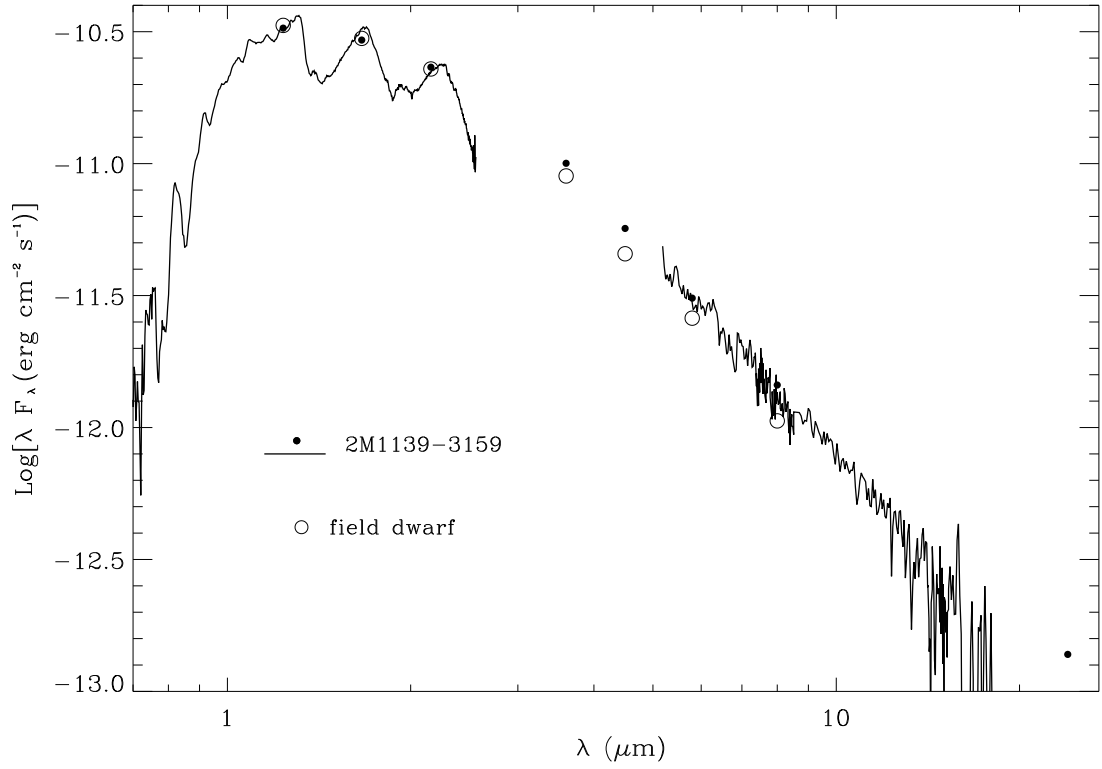}
\caption{
SEDs of the young brown dwarf 2M~1139-3159 ({\it lines and filled circles}) 
and the field dwarf VB~10 ({\it open circles}). 
The SED of 2M~1139-3159 is consistent with that of the field dwarf and 
does not show excess emission at long wavelengths from a circumstellar 
disk. The spectra of 2M~1139-3159 were obtained with SpeX and IRS in this work.
The photometric measurements are from 2MASS, \citet{pat06}, and \citet{ria06}.
The SED for VB~10 has been scaled to match that of 2M~1139-3159 at $J$, $H$,
and $K_s$. 
}
\label{fig:1139}
\end{figure}

\begin{figure}
\plotone{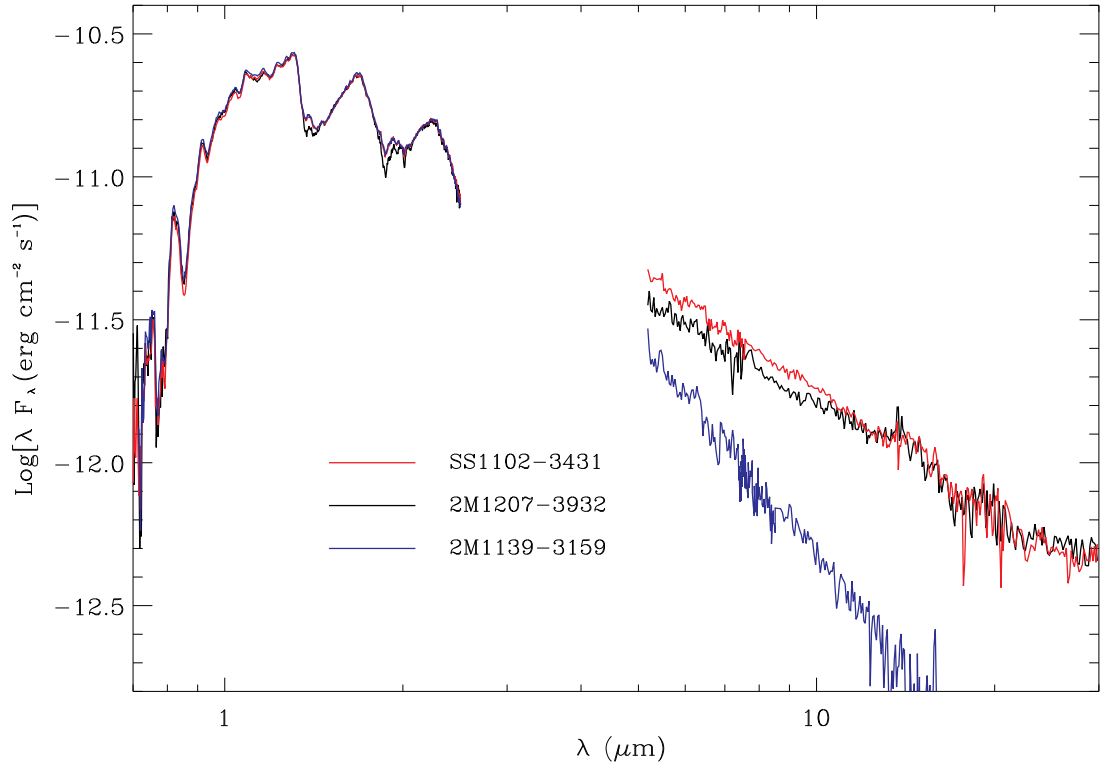}
\caption{
SEDs of the young brown dwarfs SS~1102-3431, 2M~1207-3932, and 2M~1139-3159 
({\it red, black, and blue lines}). Relative to 2M~1139-3159, 
SS~1102-3431 and 2M~1207-3932 exhibit excess emission at long wavelengths,
indicating the presence of circumstellar disks. The difference between
the SEDs of SS~1102-3431 and 2M~1207-3932 can be explained as a difference
in the heights of their inner disk walls (Figs.~\ref{fig:1102} 
and \ref{fig:1207}).
The SEDs of SS~1102-3431 and 2M~1139-3159 have been scaled to match that
of 2M~1207-3932 at 1-2.5~\micron.}
\label{fig:3bds}
\end{figure}

In Figure~\ref{fig:1139}, we plot the SED of 2M~1139-3159
from 0.7 to 24~\micron\ using photometry from 2MASS and {\it Spitzer} 
\citep{ria06} and the spectra that we obtained with SpeX and IRS. 
For comparison, we include the SED of a field dwarf with a similar spectral 
type (VB~10, M8V) after scaling it to match 2M~1139-3159 at $J$, $H$, and $K_s$.
2M~1139-3159 is slightly brighter that the field dwarf beyond 3~\micron,
but this is probably a reflection of a small difference in the
photospheric near- to mid-IR colors between pre-main-sequence objects and 
field dwarfs rather than excess emission from dust given that the mid-IR
slopes are the same for the two objects. 
Therefore, we believe that the SED of 2M~1139-3159 represents a good estimate 
of the SEDs of the stellar photospheres of the other two brown dwarfs,
2M~1207-3932 and SS~1102-3431. We compare 
the SpeX and IRS spectra of these three objects in Figure~\ref{fig:3bds}.
Relative to the photospheric SED of 2M~1139-3159, the SEDs of 
2M~1207-3932 and SS~1102-3431 exhibit significant excess emission 
longward of 5~\micron. 
These results for 2M~1207-3932 and 2M~1139-3159 are consistent with those
of previous studies, which have found evidence of a disk for
2M~1207-3932 but not for 2M~1139-3159 based on {\it Spitzer} photometry 
\citep{ria06}, ground-based photometry \citep{jay03,ste04}, and 
H$\alpha$ emission \citep{moh03,moh05}.
For SS~1102-3431, the only previous constraint on the presence of a disk
is the H$\alpha$ spectroscopy by \citet{sch05b}, who concluded that 
accretion is occurring at a very low level, if at all. 
Thus, our IRS observations provide the first definitive detection
of a disk around this object. 

In addition to detecting the presence of excess emission, previous 
mid-IR photometric measurements of 2M~1207-3932 have been used to 
constrain the strength of the silicate emission feature at 10~\micron. 
Based on ground-based photometry at 8.7 and 10.4~\micron, \citet{ste04} 
found that 2M~1207-3932 did not exhibit strong silicate emission. 
However, 2M~1207-3932 was slightly brighter at 10.4~\micron\ than 
at 8.7~\micron, which was interpreted as evidence of modest silicate emission 
by \citet{ria07}. As shown in Figure~\ref{fig:3bds}, silicate emission 
at 10 and 20~\micron\ is absent in the IRS spectrum of 2M~1207-3932.
The same is true for the other disk-bearing brown dwarf in our sample, 
SS~1102-3431. In contrast to these brown dwarf disks, the disks around 
stars in TWA do produce silicate emission \citep{uch04,fur07}.

Although silicate emission at 10 and 20~\micron\ is not present in the IRS
data for 2M~1207-3932 and SS~1102-3431, it appears that a weak emission feature
may be detected near 14~\micron\ in the spectra of both objects.
Because of its large width, the feature 
is probably produced by a mineral rather than a gas.
We cannot identify a plausible source of this emission that
is consistent with the available data for these disks (e.g., large grain
sizes, absence of other emission lines).

\subsection{Disk Model}
\label{sec:model}

We have modeled the mid-IR excess emission from 2M~1207-3932 and SS~1102-3431 
as arising from irradiated accretion disks by following the
procedures from \citet{dal98,dal99,dal01,dal06}.
For the stellar photosphere of each object, we have adopted an effective 
temperature of 2555~K, which is estimated by combining a spectral type of M8.5 
(\S~\ref{sec:spex}) with the temperature scale from \citet{luh03}.
A brown dwarf with this temperature at the age of the TW~Hya association
(10~Myr) is predicted to have a bolometric luminosity of 0.0024~$L_\odot$
and a mass of 0.025~$M_\odot$ by the evolutionary models by \citet{cha00}.
These values of temperature and 
luminosity correspond to a stellar radius of 0.25~$R_\odot$. 
For the disk calculations, we adopted a uniform accretion rate of
$\dot{M}=10^{-11}$~$\rm M_{\odot}\,yr^{-1}$ \citep{sch05b}. 
We include dust settling in the disk following the methods of \citet{dal06}.
In short, two populations of grains co-exist in the disk, both with size 
distributions given by $n(a)\sim a^{-3.5}$ where $a$ is the grain radius 
\citep{mrn}. The minimum radius for both populations is 0.005~\micron; 
grains around the midplane have a maximum radius of $a_{max}=1$~mm, while
$a_{max}$ in the upper layers is adjusted to produce the best fit to the SED.
In addition, the dust-to-gas mass ratio of the small grains is 
parameterized in terms of $\epsilon$, which is the ratio normalized by 
the standard interstellar value of $\sim0.01$. 
The model includes an inner disk wall illuminated by the central brown dwarf 
and located at the dust sublimation radius of 3.3~R$_{\star}$.
The outer radii of the disks around 2M~1207-3932 and SS~1102-3431
are not constrained by the available data; we adopt a value of $R_{out}=50$~AU
for each disk.

\begin{figure}
\plotone{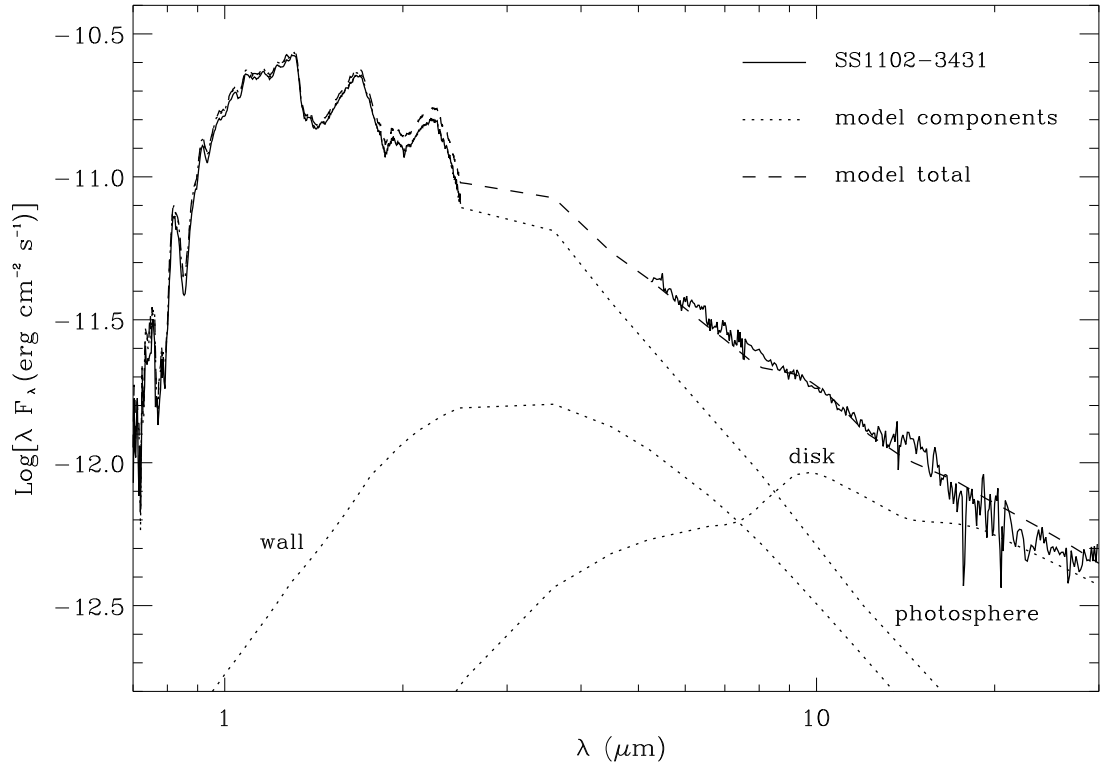}
\caption{
SED of SS~1102-3431 compared to a model for its circumstellar disk
($\alpha=0.001$, $\epsilon=0.001$, $i=60\arcdeg$, 
$\dot{M}=10^{-11}$~$\rm M_{\odot}\,yr^{-1}$, $z_{wall}=2$~$H$,
$R_{in}=3.3 R_*=0.83$~AU, $R_{out}=50$~AU, $a_{max}=5$~\micron\ for small
grains in upper disk).
We adopted the SED of 2M~1139-3159 to represent the stellar 
photosphere of SS~1102-3431.}
\label{fig:1102}
\end{figure}

\begin{figure}
\plotone{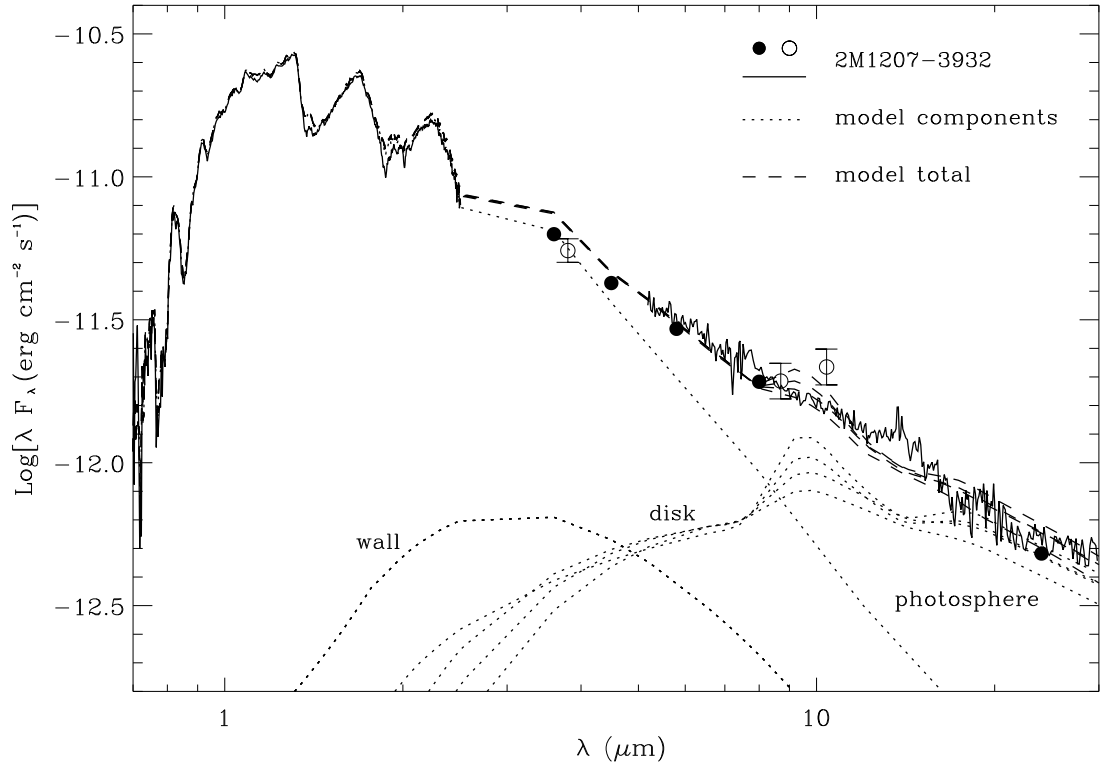}
\caption{
SED of 2M~1207-3932 compared to a model for its circumstellar disk.
The model parameters are the same as for SS~1102-3431 in Figure~\ref{fig:1102}
except that $z_{wall}=1$~$H$ and four values of $a_{max}$ 
are shown (0.25, 1, 5, and 100~\micron). The predicted strength of the silicate 
emission near 10~\micron\ is weaker for larger values of $a_{max}$. 
We adopted the SED of 2M~1139-3159 to represent the stellar 
photosphere of 2M~1207-3932. The photometric measurements are from 
\citet[][{\it open circles}]{ste04} and \citet[][{\it filled circles}]{ria06}.}
\label{fig:1207}
\end{figure}

We have calculated SEDs for a range of values of
the viscosity parameter ($\alpha$), $\epsilon$, disk inclination ($i$), and 
wall height ($z_{wall}$) and compared the results to the observed SEDs
of 2M~1207-3932 and SS~1102-3431. The SEDs of the best model fits are 
shown in Figs.~\ref{fig:1102} and \ref{fig:1207}.  The data for both objects
are well-matched by $\alpha=0.001$, $\epsilon=0.001$, and $i=60\arcdeg$.
We are able to reproduce the small difference in the SEDs of the two objects 
(Figure~\ref{fig:3bds}) by using $z_{wall}=1$~$H$ for 2M~1207-3932 and
$z_{wall}=2$~$H$ for SS~1102-3431, where $H$ is the disk scale height,
which is $1.15\times10^{-4}$~AU for the adopted sublimation temperature 
of 1400~K. These differences in $z_{wall}$ may indicate that the inner
disks have different degrees of dust settling.
We note that other combinations of values for $\alpha$, $\epsilon$, 
and $i$ also produce reasonable fits to the observed SEDs. 
Additional data, such as photometry at longer wavelengths, are needed
to further constrain these parameters. 
However, the relatively blue slopes of the mid-IR SEDs of 2M~1207-3932 and 
SS~1102-3431 definitely indicate a large degree of dust settling to the disk 
midplane \citep[i.e., small $\epsilon$ in our treatment,][]{wat08}.

The absence of silicate emission (Figure~\ref{fig:3bds}) indicates that grains 
have grown significantly in the upper disk layers \citep[cf.][]{dal06}. 
We show in Figure~\ref{fig:1207} a series of disk models in which we have
varied $a_{max}$ in the upper disk layers from 0.25~\micron\ (ISM grains) 
to 100~\micron. As expected, the silicate emission
becomes weaker as $a_{max}$ grows. Based on the comparison to our models,
the absence of silicate emission in the IRS data for 2M~1207-3932 and
SS~1102-3431 indicates that grains in the upper disk layers 
have grown to sizes of $a_{max}\gtrsim5$~\micron.

\section{Conclusions}

We have presented near- and mid-IR spectroscopy for three young brown
dwarfs in the TW~Hya association. Two of these objects, 2M~1207-3932 and
SS~1102-3431, exhibit significant mid-IR emission above that expected
from a stellar photosphere.
We have successfully reproduced the excess emission from each brown dwarf 
with an irradiated accretion disk model that includes dust settling.
Both disks exhibit high degrees of dust settling to the midplane based on
the relatively blue mid-IR slopes of their SEDs. 
In addition, our IRS spectra reveal an absence of silicate
emission at 10 and 20~\micron\ in both objects, which indicates that the 
disks have experienced significant grain growth in their upper layers 
($a_{max}\gtrsim5$~\micron).   
These results for 10-Myr-old brown dwarfs support and extend the 
previously observed trend of decreasing silicate emission with lower 
stellar masses and older ages. 
This trend may indicate that grain growth and planetesimal formation 
occur more rapidly in disks around brown dwarfs than in disks around stars,
or that grains grow faster at smaller disk radii, and it is at smaller radii
where mid-IR emission is produced for objects at lower masses 
\citep{kes07,sic07}.

\acknowledgements
We acknowledge support from grant AST-0544588 from the National Science
Foundation (A. M., K. L.), grant 172854 from CONACyT (L. A.),
grants from CONACyT and PAPIIT/DGAPA, M\'exico (P. D.), NASA
grants NAG5-9670 and NAG5-13210 (N. C., L. H.),
and NASA grants to the Spitzer-IRS instrument team at Rochester and Cornell
(W. F., B. S., D. W., C. B.).

\clearpage

\end{document}